\documentclass[letterpaper, paper,11pt]{AAS}		% for final proceedings (20-page limit)
\usepackage{bm}
\usepackage{amsmath}
\usepackage{mdframed}
\usepackage{avs}
\usepackage{amssymb}
\usepackage{algorithm}
\usepackage{algpseudocode}
\usepackage{subfig}
\usepackage[colorlinks=true, pdfstartview=FitV, linkcolor=black, citecolor= black, urlcolor= black]{hyperref}
\usepackage{footnpag}
\usepackage[superscript]{cite}
\usepackage{indentfirst}
 \usepackage[normalem]{ulem}
 \useunder{\uline}{\ul}{}
\renewcommand{\abstractname}{}

\usepackage[dvipsnames]{xcolor}

\title{\textbf{Chance-constrained, Drift-safe Guidance for Spacecraft Rendezvous}}%\textbf{Some RPOD Stuff We Did}}
\author{Andrew W. Berning Jr.\thanks{GNC Engineer, Blue Origin, ABerning@blueorigin.com}, Ethan R. Burnett\thanks{GNC Engineer, Blue Origin, EBurnett@blueorigin.com}, \ and Stefan Bieniawski\thanks{Sr Technical Fellow, Blue Origin SBieniawski@blueorigin.com}}
\date{} % Activate to display a given date or no date (if empty),
\PaperNumber{23-115}

\begin{document}
\maketitle

\begin{abstract}
	
A robust drift-safe rendezvous trajectory optimization tool is developed in this work, with applications to orbital rendezvous and proximity operations. The method is based on direct collocation and utilizes a sequential convex programming framework, and is extended from previous work to include passive safety constraints. The tool is then paired with a dispersion analysis framework to allow trajectories to be optimized subject to plant, navigation, and actuator uncertainties. The timing, direction, and magnitude of orbital maneuvers are optimized subject to the expected propellant usage, for a given navigation system performance. Representative trajectories are presented for the LEO flight regime, but the approach can also be applied to GEO and NRHO with minimal modification. 
\end{abstract} 

\section{Introduction}
Collision-free operation is a critical and driving consideration for spacecraft rendezvous guidance. In addition to the requirement that a nominal trajectory respects certain geometric safety constraints, it is also necessary to enforce safety of the ``free-drift'' trajectories, which are the trajectories that depart from the nominal trajectory in the event of any missed maneuvers. See Reference~\citenum{Breger_SafeTraj} for a discussion of drift-safe rendezvous. Safety is historically enforced in a range-based manner by one or more target-centered keep-out spheres, which are not to be violated by the free-drift trajectory for some minimum amount of time, often 24 hours.\cite{IRSIS_stand} These safety concerns compete and must coexist with other constraints and considerations, such as travel time, timing between thruster firings, and additional geometric keep-in or keep-out constraints, depending on the operational context and the phase of flight. The ideal nominal trajectory will satisfy these geometric and temporal constraints in either a time-optimal (subject to maximum allowable fuel) or fuel-optimal (subject to fixed travel time) manner. Additionally, the effects of dispersions in initial conditions, thruster error, and imperfect navigation knowledge compound to produce a distribution of possible real-world outcomes for any nominal trajectory. A trajectory that is satisfactory for flight must also be robustly achievable by the GNC system when the effects of these dispersions are included. 

Convex optimization methods are attractive for safe spacecraft rendezvous and similar challenging spaceflight guidance problems because of their inherent flexibility, computational efficiency, and well-posedness.\cite{boyd2004convex} The many competing constraints inherent to emerging space systems demand optimization formulations that can be easily modified to augment additional considerations -- a property that is readily satisfied by direct convex optimization methods. Additionally, non-convexities in the problem constraints and dynamics can be addressed via careful problem reformulation, successive convexification, and other strategies as needed. A few prior works are notable for their application of convex optimization to spaceflight guidance \cite{Benedikter2019Convex,lu2013autonomous}. Reference~\citenum{szmuk2016successive} presents a successive convexification approach for spacecraft landing, which notably addresses the challenges of the non-convexities of atmospheric flight. Reference~\citenum{berning2020suboptimal} introduces a nonlinear model-predictive control (MPC) strategy for orbit tracking in the vicinity of a near-rectilinear halo orbit (NRHO), leveraging successive quadratic programming. Reference~\citenum{BurnettSchaub_jgcd22} presents a convenient dynamic formulation and convex guidance scheme for relative motion valid in circular, elliptic, and NRHO contexts. That work poses the delta-V optimal rendezvous problem as a small second-order cone program (SOCP). Such an approach, while efficient, is not easily extended to accommodate additional constraints.

Robust safe spacecraft rendezvous guidance has previously been achieved by a two-step manual process. First, waypoints and corresponding arrival times are manually set to enforce particular desired geometric properties including drift safety. Second, dispersion analysis applies the expected navigation and control system performance in addition to perturbations on arrival states and thruster operation. In the dispersion analysis, successful closure derives minimum admissible performance capabilities of the GNC system (particularly navigation), and unsuccessful closure warrants a return to the preliminary trajectory design. Reference~\citenum{WoffindenDCrpod} showcases such an approach for trajectory design for the Dream Chaser vehicle in low-Earth orbit (LEO), and Reference~\citenum{MandEtAl_HaloRPO} is an analogous work for NRHO. These works make use of a linear covariance framework\cite{GellerLinCov} to facilitate dispersion analysis without resorting to computationally expensive Monte Carlo methods. The two-step manual design process is labor-intensive for rendezvous design with multiple competing constraints -- presenting a strong motivation for a robust chance-constrained convex optimization framework which nearly fully automates the rendezvous trajectory design process. Chance-constraint methods enable statistical assurance of rendezvous safety, and are well-suited for convex optimization methods, demonstrated by References~\citenum{berning2020spacecraft} and \citenum{berning2020control} respectively.

This work presents a novel approach for drift-safe fuel- or time-optimal robust spacecraft rendezvous guidance leveraging convex optimization to develop satisfactory safe and efficient trajectories, whose safety is additionally enforced in a chance-constrained framework accounting for the dominant dispersive effects at work in rendezvous. The structure of the paper is as follows. We begin by introducing the drift-safe rendezvous problem, then discuss our closed-loop uncertainty quantification (UQ) approach which characterizes the effects of dominant dispersions in real-world operation. Afterwards we present the successive convex optimization strategy used, which incorporates the closed-loop UQ scheme as part of the solution strategy. We then provide representative numerical results for scenarios in LEO, with accompanying discussion.

Organization and notation are borrowed from Reference~\citenum{szmuk2016successive}. Symbols in \textbf{bold} are vector quantities. Square brackets denote matrices or frame-resolved vectors, and sometimes independent variables/indices.

% References to add:
% \begin{itemize}
%     \item Woffinden's work, see e.g. References ~\citenum{WoffindenDCrpod}, \citenum{MandEtAl_HaloRPO} $\rightarrow$ Done!
% 	\item Nav error, Reference \citenum{Zanetti_OrionEFTNav} (Add to UQ section)
%     \item Successive convexification reference: \citenum{szmuk2016successive} $\rightarrow$ Done!
% 	\item SQP rendezvous reference: \citenum{berning2020suboptimal} $\rightarrow$ Done!
% 	\item Chance-constrained proximity operations reference: \citenum{berning2020spacecraft} $\rightarrow$ Done!
% 	\item Chance-constrained rendezvous reference (stochastic extension of SQP rendezvous reference): \citenum{berning2020control}, Chapter 5 $\rightarrow$ Done!
% 	\item Passive safety reference; includes``ghost trajectory" strategy that we're using: \citenum{Breger_SafeTraj} $\rightarrow$ Done!
% \end{itemize}

\section{Problem Description}

This work explores the optimal path planning of spacecraft rendezvous and proximity operations (RPO) subject to passive safety constraints. For our purposes, this means minimizing either total propellant usage or total time of flight by optimizing over the number, timing, direction, and magnitude of impulsive burn maneuvers. Additionally, we define a keep-out sphere (KOS) around the target spacecraft that the chaser spacecraft must never enter, even if control is lost and the spacecraft were to drift subject to its natural orbital motion. 

For the purposes of generating RPO trajectories it is natural to adopt the linearized Clohessy-Wiltshire (CW) equations \cite{schaub,AlfriendSFFch2}, which describe the motion of a chaser spacecraft relative to a target spacecraft orbiting a central body in a circular orbit. The linear, time-invariant properties and corresponding closed-form state transition matrix aid greatly in the numerical optimization efforts. The three-dimensional, continuous-time CW equations with plant and control matrices $A_3$ and $B_3$ are as follows:

\begin{align}
	\dot{\pmb{x}}(t) = & ~A_3 \pmb{x}(t) +B_3 \pmb{u}(t) \\
	A_3 = &~\begin{bmatrix}
		0 & 0 & 0 & 1 &0&0 \\
		0 & 0 & 0 & 0 &1&0 \\ 	
		0 & 0 & 0 & 0 &0&1 \\ 	
		3n^2 &0&0&0&2n&0 \\
		0&0&0&-2n&0&0 \\
		0&0&-n^2&0&0&0
	\end{bmatrix}
	, \\
	B_3 =  &~\begin{bmatrix}
		0_{3,3} \\
		\mathbb{I}_3
	\end{bmatrix},
\end{align}

\noindent where $\pmb{x}(t) = \begin{bmatrix}
	x_1 & x_2 & x_3 & \dot{x}_1 & \dot{x}_2 & \dot{x}_3
\end{bmatrix}^\intercal$, $\pmb{u}(t)$ is the control that represents impulsive thrust maneuvers, and $n$ is the mean motion for the circular orbit under consideration. The $x_1$ axis is along the radial direction from the central body to the target spacecraft, the $x_3$ axis is along its orbital angular momentum vector, and the $x_2$ axis completes the right-handed reference frame. The three-dimensional CW equations are used in the Uncertainty Quantification section due to the formulation of the maneuver execution error model, but the trajectory optimization model used in the Optimization section utilizes the longitudinal flight model given below: % ERB: I removed section links because they are not resolved in the PDF at all, i.e. "Section~\ref{sec:UQ} lorem ipsum" reads as "Section  lorem ipsum"

\begin{align}
	A = &~\begin{bmatrix}
		0 & 0 & 1 &0 \\
		0 & 0  & 0 &1 \\ 	
		3n^2 &0&0&2n \\
		0&0&-2n&0 \\
	\end{bmatrix}
	, \\
	B =  &~\begin{bmatrix}
		0_{2,2} \\
		\mathbb{I}_2
	\end{bmatrix},
\end{align}

\noindent with corresponding state transition matrix: 
\begin{align}
	\Phi(\Delta t) = \begin{bmatrix}
		4-3 \cos(n \Delta t) & 0 & \frac{1}{n} \sin (n \Delta t) & \frac{2}{n}(1-\cos(n \Delta t)) \\
		6(\sin(n \Delta t)-n \Delta t) & 1 & \frac{-2}{n}(1-\cos(n \Delta t)) & \frac{1}{n}(4 \sin(n \Delta t)-3n \Delta t) \\
		3n \sin(n \Delta t) & 0 & \cos(n \Delta t) & 2 \sin(n \Delta t) \\
		-6n(1-\cos(n \Delta t)) & 0 & -2n \sin (n \Delta t) & 4 \cos (n \Delta t) - 3
	\end{bmatrix} \label{eq:STM}
\end{align}

\section{Closed-Loop Uncertainty Quantification} \label{sec:UQ}

\subsection{Error and Dispersion Modeling}
For a given trajectory, it is of interest to see how the combined effects of dominant disturbances and dispersions will affect the expected distribution of states at various critical points during the rendezvous procedure. We consider the effects of vehicle delivery error (initial condition dispersions) and maneuver errors. In practice, these errors are sufficiently large to necessitate closed-loop corrections in the control of the vehicle to follow the nominal trajectory. Because of this, we also must consider the effects of navigation error on the overall closed-loop performance. Classically, such an uncertainty quantification (UQ) representation is useful for evaluating the safety and performance of nominal guidance solutions. In our case, the UQ representation is used by the convex optimization scheme to ensure that the resulting rendezvous guidance trajectory is robust to expected dispersions. In this section, we discuss our closed-loop error and dispersion representation.

Before discussing the error models used, we briefly introduce the concept of an exponentially correlated random variable (ECRV), which provides a convenient mechanism for simulating stochastic variables across a range of characteristic behaviors from constant bias to random noise.\cite{WoffindenDCrpod} For example, consider a 6-dimensional representation $\bm{z} \in \mathbb{R}^{6}$ as below.
\begin{equation}
\label{ecrv1}
\bm{z}_{k+1} = \bm{z}_{k}e^{-\Delta t/\tau} + \bm{v}_{k}
\end{equation}
where $\Delta t = t_{k} - t_{k-1}$ and $\bm{v}_{k}$ is zero-mean and normally distributed with covariance $Q_{k}$ to enable a variance of unity:
\begin{equation}
\label{ecrv2}
Q_{k} = I_{6\times 6}(1 - e^{-\frac{2\Delta t}{\tau}})
\end{equation}
The time constant $\tau$ determines the deterministicity of the ECRV. An ECRV with $\tau = 0$ behaves as white noise, and as $\tau \rightarrow \infty$, the ECRV behaves as a constant bias. The intermediate values of $\tau$ yield a pseudo-random walk. The ECRV is a versatile tool that can be used to represent navigation error or to represent any stochastic model quantities such as maneuver error parameters. Note that the ECRV equations are insensitive to the size of $\Delta t$. For a given value of $\tau$, we can simulate the ECRV for any step size of $\Delta t$ without adversely affecting the degree of stochasticity of the signal from step to step. This is especially convenient for simulation architectures with variable step sizes.

In a Monte Carlo study of dispersion, or in other applications where a computationally lean representation of onboard navigation is advantageous, the ECRV facilitates a ``stochastic nav'' representation of navigation error. The instantaneous mapping from an ECRV (with the unity-variance representation given above) to simulated navigation state error is given below.
\begin{equation}
\bm{e}_{\text{nav},k} = \sqrt{P_{\hat{\bm{x}}}(t_{k})}\bm{z}_{k}
\end{equation}
where $\bm{e}_{\text{nav},k} \in \mathbb{R}^{6}$ is the nav error at time $t_{k}$, and $P_{\hat{\bm{x}}}(t_{k}) \in \mathbb{R}^{6\times6}$ is the nav error covariance at time $t_{k}$. With this representation, for a sampling of many simulated navigation errors, the errors obey the expected statistics of the (time-varying) navigation covariance matrix. We can then specify the time-varying performance of the navigation system via the covariance history, and simulate the resulting navigation error effects without introducing a filter. The generation of nav covariance data can be achieved either by separately running a filter for a given nominal trajectory, or by geometric arguments. See for example Reference~\citenum{GeoFactor}.

To model maneuver error, we follow the Gates model, which is appropriate for our 3DOF considerations.\cite{gates_man_err} The Gates model parameterizes maneuver error in terms of four stochastic dispersions added on to the nominal delta-V vector as below:
\begin{equation}
\label{gates_ManErr1}
\Delta\bm{v}' = \Delta\bm{v} + \left(\bm{e}_{s} + \bm{e}_{p} + \bm{e}_{r} + \bm{e}_{a}\right)
\end{equation}
\begin{subequations}
	\label{gates_ManErr2}
	\begin{align}
		\bm{e}_{s} & = s\Delta\bm{v}, \ s \sim N(0,\sigma_{s}) \\
		\bm{e}_{p} & = \bm{u} \times \Delta\bm{v}, \ u_{i} \sim N(0, \sigma_{p}), i = 1,2,3 \\
		\bm{e}_{r} & = r\frac{\Delta\bm{v}}{\Delta v}, \ r \sim N(0, \sigma_{r}) \\
		\bm{e}_{a} & = \bm{w} \times \frac{\Delta\bm{v}}{\Delta v}, \ w_{i} \sim N(0, \sigma_{a}), i = 1,2,3
	\end{align}
\end{subequations}
where $\Delta \bm{v}'$ is the erroneous delta-V vector, the nominal delta-V vector is unprimed, and $N(0,\sigma)$ indicates a zero-mean normally distributed parameter with standard deviation $\sigma$. The term $\bm{e}_{s}$ is the proportional magnitude error, $\bm{e}_{p}$ is the proportional pointing error, $\bm{e}_{s}$ is the fixed magnitude error, and $\bm{e}_{a}$ is the fixed pointing error. The errors can be uncorrelated from burn to burn, or in a more realistic representation, they could be represented by ECRVs. The maneuver error can also be defined in a ``principal-error" coordinate frame $\{\bm{e}_{1},\bm{e}_{2},\bm{e}_{3}\}$ whereby the $\bm{e}_{1}$ axis is parallel to the nominal delta-V vector, and the $\bm{e}_{2}$ and $\bm{e}_{3}$ axes are perpendicular and arbitrarily oriented. The above maneuver error definition has a statistically equivalent formulation with covariance defined as $P_{G} = \text{diag}(\sigma_{r}^{2} + \Delta v^{2}\sigma_{s}^{2}, \sigma_{a}^{2} + \Delta v^{2}\sigma_{p}^{2}, \sigma_{a}^{2} + \Delta v^{2}\sigma_{p}^{2})$ in the principal-error coordinates -- see e.g. Reference~\citenum{Wagner_Cassini}. An alternate maneuver error model given in Reference~\citenum{GellerLinCov} is also popular, incorporating the effects of attitude error. Setting the attitude error to zero, and using the same dispersion statistics for both the fixed magnitude and fixed pointing errors, it is possible to make these error models equivalent.

To simulate the effects of dispersions on a closed-loop control scheme, the only data needed are the dispersion/error statistics, nominal burn times and delta-V vectors, and the nominal state at each burn time. Our UQ formulation simulates a sample of dispersed trajectories, using the STM to propagate the sampled states between maneuver times. To avoid the computational burdens of a Monte Carlo representation, we use a linear covariance representation to recover the statistically expected spread of the dispersed trajectories between maneuver times, and also for the free-drift trajectories in the event of any missed maneuvers. This procedure is used to verify that the dispersed trajectories do not intersect the keep-out sphere, to some degree of specified confidence, e.g. 3-sigma. The dispersed state covariance $P_{\bm{x}}$ is propagated using the STM as below:
\begin{equation}
\label{cov_state1}
P_{\bm{x}}(t_{k+1}) = \Phi(t_{k+1},t_{k})P_{\bm{x}}(t_{k})\Phi^{\top}(t_{k+1},t_{k})
\end{equation}

\subsection{Closed-Loop Control Approach}
At the maneuver times, the sampled trajectories apply a maneuver that is a linearized Lambert-style correction of the nominal maneuver, plus the maneuver errors following the Gates model formulation. The linearized Lambert-style correction computes at the $j$\textsuperscript{th} burn time $t_{j}$ the expected corrected delta-V (based on the current nav state) needed to return the trajectory back to the nominal trajectory at burn time $t_{j+1}$. In this manner, corrective maneuvers are only applied at the nominal burn times. This linearized two-burn corrective Lambert scheme (see e.g. Reference~\citenum{burnett2021phd}, Chapter 7) is given using the following equations, where only the first maneuver is applied at a waypoint, and as a consequence the final waypoint (final maneuver) can only correct the final velocity state:
\begin{subequations}
	\label{Lambert1}
	\begin{align}
		\Phi(t_{j+1},t_{j}) & = \begin{bmatrix}
			\Phi_{rr} & \Phi_{rv} \\ \Phi_{vr} & \Phi_{vv}
		\end{bmatrix} \\
		t_{j}: \ \Delta\bm{v}_{1} & = \Phi_{rv}^{-1}\left(\bm{r}^{*}(t_{j+1}) - \Phi_{rr}\hat{\bm{r}}(t_{j})\right) - \hat{\bm{v}}(t_{j}^{-}) \\
		\hat{\bm{v}}(t_{j}^{+}) & = \hat{\bm{v}}(t_{j}^{-}) + \Delta\bm{v}_{1} \\
		t_{j+1}: \ \Delta\bm{v}_{2} & = \bm{v}^{*}(t_{j+1}) - \left(\Phi_{vr}\hat{\bm{r}}(t_{j}) + \Phi_{vv}\hat{\bm{v}}(t_{j}^{+}) \right)
	\end{align}
\end{subequations}
where $\hat{\bm{r}}$ and $\hat{\bm{v}}$ denote the nav-estimated target-relative position and velocity, and $t_{j}^{-}$ and $t_{j}^{+}$ denote time $t_{j}$ in the instants before and after the $j$\textsuperscript{th} maneuver. The equation for $\Delta\bm{v}_{2}$ is only a projected estimate for the next maneuver, i.e. $\Delta\bm{v}_{1}(t_{j+1})$, and is never applied. We sample enough trajectories at the burn times to ensure that the statistical outcome of the erroneous maneuvers and dispersed states is recovered, and can use covariance propagation otherwise. 

For classical dispersion analysis of the nominal trajectories, we have the capability to either apply a fully Monte Carlo approach, or a nearly purely linear covariance (LinCov) approach, with the exception of the state sampling at the burn times. In the case that only the linearized relative motion dynamics are considered, the statistical result of these two approaches will agree. When the UQ scheme is implemented within the robust guidance algorithm, the UQ algorithm is made as computationally efficient as possible. For example, there is minimal use of Monte Carlo methods, and we primarily use LinCov techniques. Additionally, the ECRVs are only sampled as-needed, at the burn times. Recall that Eqs. \eqref{ecrv1} - \eqref{ecrv2} can be applied with any $\Delta t$ for a given choice of $\tau$.

\subsection{Example LEO Application}
Here, we demonstrate the closed-loop dispersion analysis functionality by example analysis of a ``double-coelliptic'' rendezvous trajectory in LEO, similar to the example given in Reference~\citenum{WoffindenDCrpod}. For this example, the target is in a circular low-Earth orbit with semimajor axis of 6738 km. The chaser starts on a coelliptic with a 4 km R-bar offset, transfers to a coelliptic with a 1.4 km R-bar offset, then transfers to a 750 m hold point along the V-bar. The waypoint states and times are given in Table~\ref{table:uq1}, and the nominal delta-V maneuvers are given in Table~\ref{table:uq2}. Relevant data for the dispersion statistics are then summarized in Table~\ref{table:uq3}.
\begin{table}[h!]
\caption{Waypoints and Timing for LEO Rendezvous Example}
	\begin{tabular}{l|l|l|l|l}
	\label{table:uq1}
	Waypoint & Type                      & Chaser State (LVLH) & Transfer Time & Hold Time \\ \hline
	CT       & Coelliptic, 4 km & $(x, y, z) = (-4, -17.5, 0)$ km    & N/A           & 0.5 min   \\
			 &							 & $(\dot{x}, \dot{y}, \dot{z}) = (0, 6.849, 0)$ m/s &				  &			  \\
	NSR      & Coelliptic, 1.4 km        & $(x, y, z) = (-1.4, -7.5, 0)$ km   & 35.5 min      & 0 min     \\
			&							 & $(\dot{x}, \dot{y}, \dot{z}) = (0, 2.397, 0)$ m/s &				  &			  \\
	AI       & Approach initiation       & $(x, y, z) = (-1.4, -0.75, 0)$ km   & 46.875 min    & 0 min     \\
			&							 & $(\dot{x}, \dot{y}, \dot{z}) = (0.741, 2.716, 0)$ m/s &				  &			  \\
	HP750    & 750 m V-bar hold    & $(x, y, z) = (0, 0.75, 0)$ km   & 36 min        & Any				\\
			&							 & $(\dot{x}, \dot{y}, \dot{z}) = (0, 0, 0)$ m/s &				  &		     
	\end{tabular}
\end{table}

\begin{table}[h!]
\caption{Maneuver Data for LEO Rendezvous Example}
	\begin{tabular}{l|l|l|l|l}
	\label{table:uq2}
	Burn Number & Time since CT & Delta-V Vector (LVLH)      & Magnitude  & Associated Waypoint \\ \hline
	1           & 0.5 min       & $(0.5415, 0.7494, 0)$ m/s  & 0.9245 m/s & Leave CT            \\
	2           & 35.5 min      & $(-0.6195, 0.7345, 0)$ m/s & 0.9609 m/s & Arrive at NSR       \\
	3           & 82.375 min    & $(0.739, 0.3187, 0)$ m/s   & 0.8048 m/s & Start AI            \\
	4           & 118.375 min   & $(0.1795, 0.4804, 0)$ m/s  & 0.5129 m/s & Arrive at HP750    
	\end{tabular}
\end{table}

\begin{table}[h!]
\caption{Dispersion Data for LEO Rendezvous Example}
	\begin{tabular}{l|l}
	\label{table:uq3}
	Dispersion Parameters & Values                                                                                                                                                                        \\ \hline
	Initial Conditions    & $P_{\bm{x}_{o}} = \begin{bmatrix} \sigma_{R}^{2}I_{3\times 3} & 0_{3\times 3} \\ 0_{3\times 3} & \sigma_{V}^{2}I_{3\times 3} \end{bmatrix}$, $\sigma_{R} = 40$ m, $\sigma_{V} = 5$ cm/s \\
	Navigation            & Initial 3-sigma RSS: $233.46$ m, $22.49$ cm/s. ECRV: $\tau = 3.6$ hrs                                                                                                                    \\
	Maneuver Errors       & $\sigma_{s} = 2\times 10^{-3}$, $\sigma_{r} = 0.3$ mm/s, $\sigma_{p} = 3\times 10^{-4}$ rad., $\sigma_{a} = 0.3$ mm/s                                                                                                           
	\end{tabular}
\end{table}

\begin{figure}[h!]
\centering
\includegraphics[]{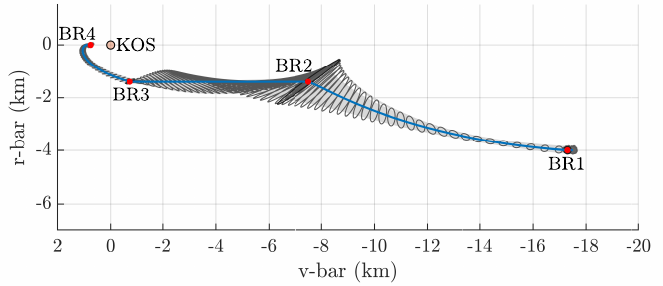}
\caption{LEO Trajectory with 3-Sigma Uncertainty Bounds}
\label{fig:uq1}
\end{figure}

\begin{figure}[h!]
\centering
\includegraphics[]{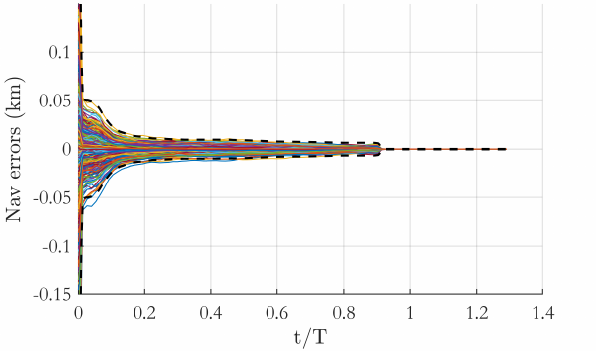}
\caption{Stochastic Navigation Error with 3-sigma RSS Limits}
\label{fig:uq2}
\end{figure}

Figure~\ref{fig:uq1} gives the resulting nominal trajectory with $3\sigma$ bounds on the dispersed trajectories. The nominal trajectory is given in blue, with all four maneuvers labeled, and the 150 m radius keep-out sphere is also labeled. A random sampling of 500 dispersed trajectories produces the trajectories in light gray. Closed-loop linearized Lambert correction facilitates a contraction of the dispersed trajectories by the time the final V-bar waypoint is reached. Free-drift trajectories are not shown in this figure. The large dispersion at BR2 is because the nav error and position dispersions were fairly high at BR1. However, note the significant contraction of the uncertainty bounds from BR2 to BR3, corresponding with greatly improved nav error by the time BR2 is reached. Figure~\ref{fig:uq2} gives the stochastic nav error (position components) in this simulation, with corresponding $3\sigma$ RSS uncertainty bounds. As expected, $\sim$99.7\% of the nav error stays within these bounds. Note the very large nav error early in the simulation, which negatively impacts the accuracy of BR1.

\begin{figure}[h!]
\centering
\includegraphics[]{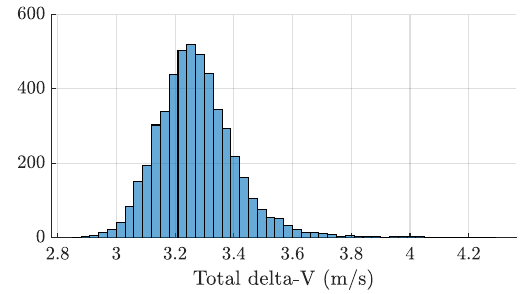}
\caption{Delta-V Distribution, 5000 Run Monte Carlo}
\label{fig:uq3}
\end{figure}

\begin{figure}[h!]
\centering
\includegraphics[]{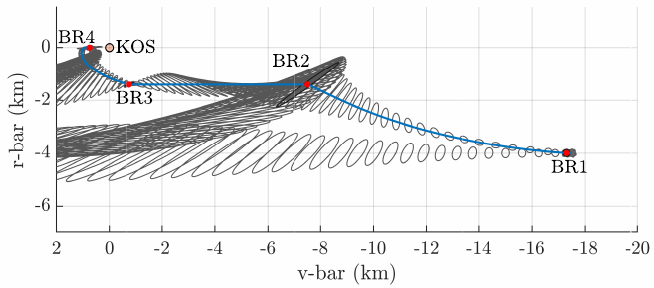}
\caption{LEO Trajectory with Free-Drift 3-Sigma Bounds}
\label{fig:uq4}
\end{figure}

The dispersion analysis formulation enables efficient study of trajectory performance. For example, for this LEO case, Figure~\ref{fig:uq3} gives the dispersion of total delta-V in a 5000 run Monte Carlo. Finally, Figure~\ref{fig:uq4} showcases the free-drift computation with linear covariance propagation. Note the large dispersion in the lower 4 km coelliptic in the event that maneuver BR1 is missed. This figure shows that this LEO trajectory is drift-safe to $3\sigma$ confidence, because none of the trajectory $3\sigma$ ellipsoids intersect the keep-out sphere. Note that this result is computed with less than 1\% of the computation time used for the Monte Carlo study that produced Figure~\ref{fig:uq1}.

\section{Passively Safe Optimization} \label{sec:optimization}

In this section, we describe the problem of optimizing burn timing, direction, and magnitude subject to certain safety constraints and the uncertainties computed with the UQ scheme presented above. For purposes of computational efficiency and ease of formulation, this optimal control problem (OCP) is formulated first as a direct collocation parameter optimization problem and then as a sequence of SOCPs to be solved by any standard interior point method solvers. This paper builds up to the final SOCP similar to how it is presented in \citenum{szmuk2016successive}: first the continuous-time OCP is presented, then the discrete-time OCP, followed by the chance-constrained OCP, and finally the successive convexification algorithm. 

\subsection{Continuous-Time Formulation}

The continuous-time problem formulation is summarized in Problem 1. The objective function minimizes either total propellant usage or time of flight while adhering to the CW dynamics and two separate safety constraints. Eq.~\eqref{eq:KOS} restricts the trajectory from entering a keep-out sphere (KOS) centered on the target spacecraft at the origin with radius $r_{\text{KOS}}$. Eq. \eqref{eq:free_drift} enforces a passive safety constraint, ensuring that if the spacecraft were to become unresponsive at any point during a mission and miss a maneuver, its free-drift natural motion would not bring it within the KOS until at least time $t_{\text{safe}}$. In the problem described below, and in problems to follow, let $\Phi_{r}(\tau)$ be the position-associated (top 2 rows) of the $4\times4$ longitudinal STM as below:
\begin{equation}
\label{phisubr}
\Phi_{r}(\tau) = [I_{2\times2} \ 0_{2\times2}]\Phi(\tau)
\end{equation}
where the longitudinal STM is given by Eq.~\eqref{eq:STM}.
\begin{mdframed}[frametitle={Problem 1: Continuous Time}]
	\allowdisplaybreaks
	\begin{align}
		\underline{\mathbf{Objective~ Function:}} & \nonumber \\
		\min_{t_f, \pmb{u}} \quad & \int_0^{t_f} ||\pmb{u}(t)|| dt \quad \text{subject to:} \\ 
		& \mathrm{OR} \nonumber \\
		\min_{t_f, \pmb{u}} \quad & t_f \quad \text{subject to:} \\ 
		\underline{\mathbf{Boundary~ Conditions:}} & \nonumber \\
		\pmb{X}(0) & = \begin{bmatrix}
			\pmb{r}_i \\
			\pmb{v}_i
		\end{bmatrix} , \pmb{X}(t_f) = \begin{bmatrix}
			\pmb{r}_f \\
			\pmb{v}_f
		\end{bmatrix}, \\
		\underline{\mathbf{Dynamics:}} &  \nonumber \\
		\dot{\pmb{X}} & = A \pmb{X} + B \pmb{u}
		\nonumber  \\
		\underline{\mathbf{Constraints:}} &  \nonumber \\
		||\pmb{r}(t)|| &\geq r_{\text{KOS}} & \forall t \in [0,t_f] \label{eq:KOS} \\
		||\Phi_{r}(\tau)\pmb{X}(t)|| &\geq r_{\text{KOS}} & \forall t \in [0,t_f] ~ \land ~ \forall \tau \in [0,t_{\text{safe}}] \label{eq:free_drift}
	\end{align}
\end{mdframed}

\subsection{Discrete-Time Formulation}

In order to express this continuous-time OCP as a parameter optimization problem, we adopt a direct collocation strategy and divide the trajectory into $N-1>0$ time segments of unequal lengths $\Delta t [k]$. Thus we are left with $N > 1$ nodes at which the trajectory must be computed. For a given $N$, $\Delta t$ can be related to $t_f$ by:

\begin{align}
	k_f \triangleq N-1 \\
	t_f = \sum_{k=0}^{k_f} \Delta t [k]
\end{align}

For the problems presented below, we will use $k$ to refer to time $t = \sum_{\kappa = 0}^k \Delta t [\kappa]$ and will solve for $\Delta t [k], k \in [0,k_f]$ as opposed to $t_f$ as in Problem 1. Assume that $k \in [0,k_f]$ unless otherwise specified.

The discretization scheme utilized in the OCPs below relies on the state transition matrix, $\Phi (\Delta t)$ defined in Eq. \eqref{eq:STM}. This is computationally advantageous as there is a closed-form solution available, and it allows the dynamics to remain physically feasible no matter the length of the discretization interval $\Delta t [k]$. The passive safety constraint is discretized by enforcing the constraint at every time in set $\mathbb{T}$:

\begin{align}
	\mathbb{T} =  \{0, \gamma , 2 \gamma, 3 \gamma, \ldots ,  t_{\text{safe}}  \},
\end{align}

\noindent where $\frac{1}{\gamma}$ is the frequency at which the passive safety constraint is enforced. Notice that Eq. \eqref{eq:disc_drift_safe} subsumes both Eqs. \eqref{eq:KOS} and \eqref{eq:free_drift}.

The discretized version of Problem 1 is presented in Problem 2 below.

\begin{mdframed}[frametitle={Problem 2: Discrete Time}]
	\allowdisplaybreaks
	\begin{align}
		\underline{\mathbf{Objective~ Function:}} & \nonumber \\
		\min_{\Delta t, \pmb{u}} \quad & \sum_{k=0}^N ||\pmb{u}[k]|| \\ 
		& \mathrm{OR} \nonumber \\
		\min_{\Delta t, \pmb{u}} \quad & \sum_{k=0}^{N-1} \Delta t [k] \\ 
		\underline{\mathbf{Boundary~ Conditions:}} & \nonumber \\
		\pmb{X}[0] & = \begin{bmatrix}
			\pmb{r}_i \\
			\pmb{v}_i
		\end{bmatrix} + B \pmb{u}[0], \pmb{X}[k_f] = \begin{bmatrix}
		\pmb{r}_f \\
		\pmb{v}_f
	\end{bmatrix}, \\
		\underline{\mathbf{Dynamics:}} &  \nonumber \\
		\pmb{X}[k+1] & = \Phi (\Delta t [k]) \pmb{X}[k] + B \pmb{u}[k+1] & \forall k \in [0,k_f) \\
		\nonumber  \\
		\underline{\mathbf{Constraints:}} &  \nonumber \\
		||\Phi_{r}(\tau)\pmb{X}[k]|| &\geq r_{\text{KOS}} & \forall \tau \in \mathbb{T} \label{eq:disc_drift_safe}
	\end{align}
\end{mdframed}

\subsection{Chance Constraints}

In order to be robust to uncertainties in the plant dynamics, actuator errors, and navigational estimates, it is desirable to enforce the KOS state constraint in a chance-constrained manner:

\begin{equation}
	\textrm{Prob}\{ \Phi_{r}(\tau)\pmb{X}[k] \geq r_{\text{KOS}}  \} \geq \beta, \forall \tau \in \mathbb{T} ,
\end{equation} 

\noindent where $\beta \in [0,1]$ is the desired probability of constraint satisfaction. 

In order to enforce \eqref{eq:disc_drift_safe} in this chance-constrained manner, we construct a probability ellipse \cite{malyshev1992optimization}:

\begin{align}
	\textrm{Prob}\{ \pmb{\omega} \in \mathbb{R}^2 : (\pmb{\omega} - \pmb{r}(t))^\top \Sigma_2^{-1}(t) (\pmb{\omega} - \pmb{r}(t)) \leq c^2      \} = \beta, \label{eq:prob_ellipse}
\end{align}
% ERB: The use of bold omega here is likely to make people think of angular velocity

\noindent where $\Sigma_2^{-1}(t) = P_{x,(1:2,1:2)}(t)$ is the block of the state covariance matrix $P_x (t)$ corresponding to the position states, and $c$ is solved for using the two degree-of-freedom chi-squared distribution\cite{lancaster2005chi}. Thus Eq. \eqref{eq:disc_drift_safe} may be enforced probabilistically by ensuring that the intersection between \eqref{eq:prob_ellipse} and the KOS is null, for every position state the vehicle might encounter and its corresponding covariance matrix: 

\begin{align}
		\{ \pmb{\omega} \in \mathbb{R}^2 : & (\pmb{\omega} - \Phi_{r}(\tau)\pmb{X}[k])^\top \Sigma_2^{-1}[\tau,k] (\pmb{\omega} - \Phi_{r}(\tau)\pmb{X}[k]) \leq c^2  \} ~ \cap \nonumber \\
	\{ \pmb{\omega} \in \mathbb{R}^2 : &||\pmb{\omega}|| \leq r_{\text{KOS}} \} = \emptyset & \forall \tau \in \mathbb{T}.
	 \label{eq:disc_drift_safe_set} 
\end{align}

In practice, this is achieved by adding a scalar buffer, $r_b[\tau,k]$, to the right-hand side of Eq. \eqref{eq:disc_drift_safe} of the appropriate size to ensure that the $\beta$-probability ellipse does not intersect the keep out zone of radius $r_{\text{KOS}}$. Notice that this buffer size corresponds to the covariance matrix for the state in question, which varies for each $k \in [0,k_f]$ and $\tau \in \mathbb{T}$. Determining the exact buffer $r_b[\tau,k]$ to satisfy this constraint would involve solving a quadratically-constrained quadratic program (QCQP) for every state in consideration, which could negatively impact the total computation time of the algorithm. Instead, a conservative approximation is adopted here, basing the buffer sizing on the circumscribing circle of the probability ellipse in Eq. \eqref{eq:prob_ellipse}: 

\begin{align}
	r_b[\tau,k] = \max \bigg \{\frac{1}{\sqrt{\lambda_1}} ,\frac{1}{\sqrt{\lambda_2}} \bigg \}
\end{align}

\noindent where $\lambda_1$ and $\lambda_2$ are the eigenvalues of $\Sigma_2^{-1}[\tau,k]$. 

This chance constraint is reflected in Problem 3 below.

\begin{mdframed}[frametitle={Problem 3: Chance-Constrained}]
	\allowdisplaybreaks
	\begin{align}
	\underline{\mathbf{Objective~ Function:}} & \nonumber \\
	\min_{\Delta t, \pmb{u}} \quad & \sum_{k=0}^N ||\pmb{u}[k]|| \\ 
	& \mathrm{OR} \nonumber \\
	\min_{\Delta t, \pmb{u}} \quad & \sum_{k=0}^{N-1} \Delta t [k] \\ 
	\underline{\mathbf{Boundary~ Conditions:}} & \nonumber \\
	\pmb{X}[0] & = \begin{bmatrix}
		\pmb{r}_i \\
		\pmb{v}_i
	\end{bmatrix} + B \pmb{u}[0], \pmb{X}[k_f] = \begin{bmatrix}
		\pmb{r}_f \\
		\pmb{v}_f
	\end{bmatrix}, \\
	\underline{\mathbf{Dynamics:}} &  \nonumber \\
	\pmb{X}[k+1] & = \Phi (\Delta t [k]) \pmb{X}[k] + B \pmb{u}[k+1] & \forall k \in [0,k_f) \label{eq:dynamics} \\
	\nonumber  \\
	\underline{\mathbf{Constraints:}} &  \nonumber \\
	||\Phi_{r}(\tau)\pmb{X}[k]|| &\geq r_{\text{KOS}} + r_{b}[\tau,k] & \forall \tau \in \mathbb{T} \label{eq:disc_drift_safe_CC}
\end{align}
\end{mdframed}

\subsection{Successive Convexification}

The methods for convexifying a nonconvex optimization problem using iterative, trust region methods have been covered extensively in the literature \cite{szmuk2016successive,szmuk2020successive,casoliva2013spacecraft,liu2014solving,carson2006model} but will be briefly re-stated here for readability.

Eqs. \eqref{eq:dynamics} and \eqref{eq:disc_drift_safe_CC} in Problem 3 represent nonconvex constraints: the former due to the inclusion of an optimization parameter $\Delta t [k]$ directly multiplying another optimization parameter $\pmb{X}$, and the latter due to the presence of a norm exclusion constraint. These nonconvexities are addressed by solving a sequence of $n_{SC} > 1$ SOCP problems, the first of which ($i=0$) is initialized with a reasonable guess as to the optimization parameters and the remainder of which ($i>0$) are linearized about the solution to the $(i-1)^{th}$ iteration. 

The successive convexification scheme is initialized with a time of flight guess, $t_{f,0}$, and a linear position trajectory $\pmb{r}[k]$ from $\pmb{r}_i$ to $\pmb{r}_f$:

\begin{equation}
	\pmb{r}_0[k] = \pmb{r}_i + \frac{k}{k_f}(\pmb{r}_f - \pmb{r}_i) \quad \forall k \in [0,k_f] \label{eq:ref_pos}
\end{equation}

Problem 4 is then solved once using a constant discretization interval $\Delta \xi = t_{f,0} / k_f$ and linearizing Eq. \eqref{eq:disc_drift_safe_CC} about the reference trajectory \eqref{eq:ref_pos}. 

The remaining SC iterations ($i>0$) solve Problem 5 in which both the dynamics and the KOS constraint are linearized about the  $(i-1)^{th}$ iteration. For this implementation, there is no trust region on the position trajectories, and a simple linear trust region on the discretization intervals:

\begin{align}
	\Delta t [k] - \Delta t_0 [k] &\leq \phi \Delta t_0 [k] \\
	\Delta t [k] - \Delta t_0 [k] &\geq -\phi\ \Delta t_0 [k] 
\end{align}
 \noindent where $\Delta t_0 [k]$ is the $k^{th}$ discretization interval from the $(i-1)^{th}$ SC iteration and $\phi$ is the trust region parameter, here set to $\phi = 0.1$. 

This entire successive convexification process is summarized in Algorithm 1 below. 

\begin{mdframed}[frametitle={Problem 4: SCVX Initialization}]
	\allowdisplaybreaks
	\begin{align}
	\underline{\mathbf{Objective~ Function:}} & \nonumber \\
	\min_{\pmb{u}} \quad & \sum_{k=0}^N ||\pmb{u}[k]|| \\ 
	\underline{\mathbf{Boundary~ Conditions:}} & \nonumber \\
	\pmb{X}[0] & = \begin{bmatrix}
		\pmb{r}_i \\
		\pmb{v}_i
	\end{bmatrix} + B \pmb{u}[0], \pmb{X}[k_f] = \begin{bmatrix}
		\pmb{r}_f \\
		\pmb{v}_f
	\end{bmatrix}, \\
	\underline{\mathbf{Dynamics:}} &  \nonumber \\
	\pmb{X}[k+1] & = \Phi (\Delta \xi) \pmb{X}[k] + B \pmb{u}[k+1] & \forall k \in [0,k_f)  \\
	\nonumber  \\
	\underline{\mathbf{Constraints:}} &  \nonumber \\
 	r_{\text{KOS}} + r_{b}[\tau,k] & \leq 	||\Phi_{r}(\tau)\pmb{X}_0[k]|| \\ & \ \ + \frac{\partial ||\Phi_{r}(\tau)\pmb{X}[k]||}{\partial \pmb{r}} \bigg |_{\pmb{r}_0[k]} \bigg [ \Phi_{r}(\tau)\pmb{X}[k] - \Phi_{r}(\tau)\pmb{X}_0[k] \bigg] & \forall \tau \in \mathbb{T} 
\end{align}
\end{mdframed}
% Copying the old equation here because I'm not 100% sure my edits are correct: r_{\text{KOS}} + r_{b}[\tau,k] & \leq 	||\Phi (\tau)\pmb{r}_0[k]|| + \frac{\partial ||\Phi (\tau)\pmb{r}[k]||}{\partial \pmb{r}} \bigg |_{\pmb{r}_0[k]} \bigg [ \Phi (\tau)\pmb{r}[k] - \Phi (\tau)\pmb{r}_0[k] \bigg] & \forall \tau \in \mathbb{T} 

\begin{mdframed}[frametitle={Problem 5: SCVX}]
	\allowdisplaybreaks
	\begin{align}
	\underline{\mathbf{Objective~ Function:}} & \nonumber \\
	\min_{\Delta t, \pmb{u}} \quad & \sum_{k=0}^N ||\pmb{u}[k]|| \\ 
	& \mathrm{OR} \nonumber \\
	\min_{\Delta t, \pmb{u}} \quad & \sum_{k=0}^{N-1} \Delta t [k] \\ 
	\underline{\mathbf{Boundary~ Conditions:}} & \nonumber \\
	\pmb{X}[0] & = \begin{bmatrix}
		\pmb{r}_i \\
		\pmb{v}_i
	\end{bmatrix} + B \pmb{u}[0], \pmb{X}[k_f] = \begin{bmatrix}
		\pmb{r}_f \\
		\pmb{v}_f
	\end{bmatrix}, \\
	\underline{\mathbf{Dynamics:}} &  \nonumber \\
	\pmb{X}[k+1] & = \Phi (\Delta t_0 [k]) \pmb{X}[k] \\ & \ \ + \bigg [ \frac{\partial \Phi()}{\partial \Delta t} \bigg |_{\Delta t_0[k]} \bigg (\Delta t[k] - \Delta t_0[k] \bigg) \bigg]  + B \pmb{u}[k+1] & \forall k \in [0,k_f)  \\
	\nonumber  \\
	\underline{\mathbf{Constraints:}} &  \nonumber \\
 	r_{\text{KOS}} + r_{b}[\tau,k] & \leq 	||\Phi_{r}(\tau)\pmb{X}_0[k]|| \\ & \ \ + \frac{\partial ||\Phi_{r}(\tau)\pmb{X}[k]||}{\partial \pmb{r}} \bigg |_{\pmb{r}_0[k]} \bigg [ \Phi_{r}(\tau)\pmb{X}[k] - \Phi_{r}(\tau)\pmb{X}_0[k] \bigg] & \forall \tau \in \mathbb{T} 
\end{align}
\end{mdframed}

\begin{algorithm}
	\caption{Successive Convexification}\label{alg:SCVX}
	\begin{algorithmic}[1]
		\State Specify environmental parameter $n$, boundary conditions (e.g. $\pmb{r}_i$,$\pmb{v}_i$,$\pmb{r}_f$,$\pmb{v}_f$), and algorithmic parameters (e.g. $N$, $n_{SC}$, etc.)
		\State Compute initial reference trajectory from time of flight guess $t_{f,0}$
		\State Solve Problem 4 and set solution to be new reference trajectory
		\While{Convergence criteria remain unmet}
		\State Solve Problem 5 and set solution to be new reference trajectory
		\EndWhile
	\end{algorithmic}
\end{algorithm}

Lastly, a grid search is wrapped around Algorithm 1 to search through different numbers of burn maneuvers, $N \in \mathbb{Z}$, to find the optimal. 

\section{Results and Discussion}

\begin{figure}[h!]
	\centering
	\includegraphics[width=0.75\linewidth]{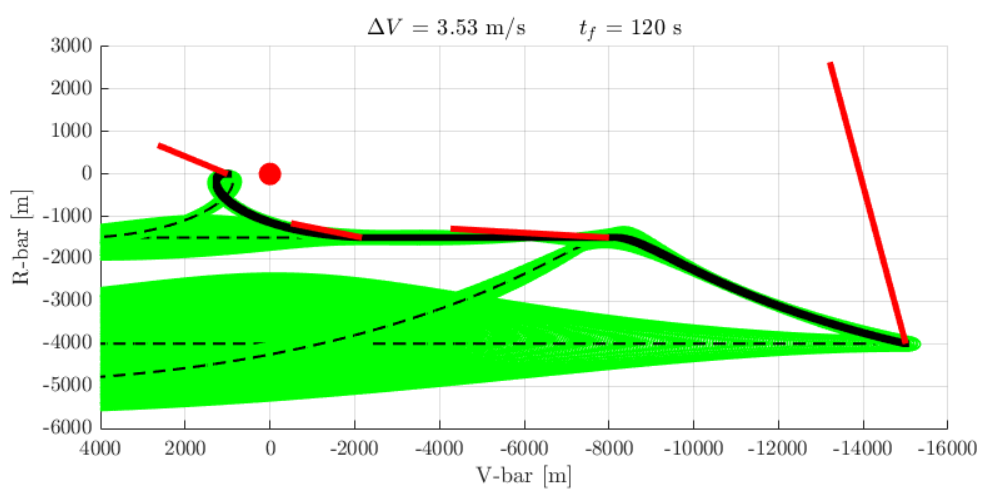}
	\caption{Prescribed trajectory}
	\label{fig:traj_prescribed}
\end{figure}

Figure \ref{fig:traj_prescribed} illustrates a control example where Algorithm 1 is solved with constraints on $\pmb{r}[k],~ k = {1,2,3,4}$ such that the position states at the initial, final, and mid-course burns are in similar positions to the example shown in \citenum{WoffindenDCrpod}. The orbital mean motion parameter $n$ is set to represent a LEO orbit and $\beta = 0.99$. The total time of flight is constrained to 120 minutes, but the time between maneuvers is allowed to be optimized for minimum propellant usage. The solid black line represents the nominal trajectory, the red lines represent both magnitude and direction of impulsive maneuvers, the dashed black lines represent the 'free drift' trajectories that would result in the case of spacecraft LOC, and the green shaded regions represent the corresponding uncertainty tubes built from the uncertainty ellipses defined in Eq. \eqref{eq:prob_ellipse}. The resultant total $\Delta V$ of $3.53$ m/s will serve as a control against which to measure the results of Algorithm with both the minimum time and minimum propellant objective functions. 

\begin{figure}[h!]
	\centering
	\includegraphics[width=0.75\linewidth]{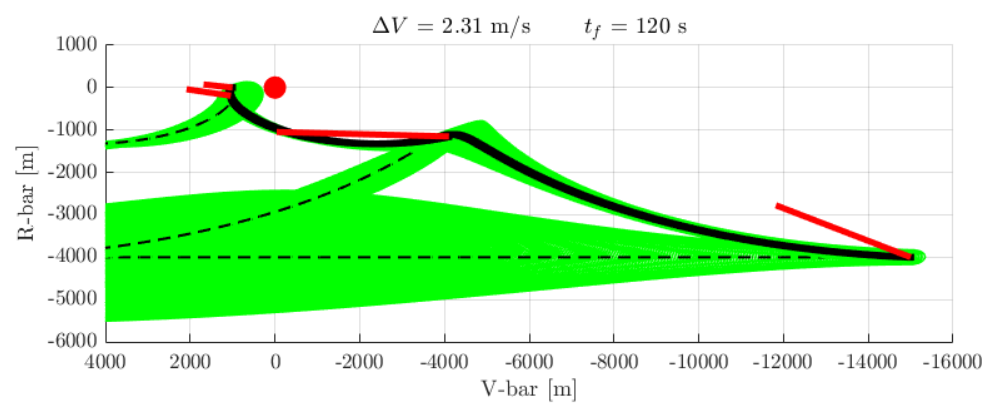}
	\caption{Minimum-propellant trajectory}
	\label{fig:traj_min_prop}
\end{figure}

In Figure \ref{fig:traj_min_prop}, all parameters remain the same except that the constraints on $\pmb{r}[2]$ and $\pmb{r}[3]$ are removed and Algorithm 1 is solved subject to the minimum propellant objective function and the total time of flight is constrained to be less than 120 minutes. Notice how the second maneuver is shifted later into the trajectory, and the last two burns are timed such that the free-drift uncertainty tube from the penultimate burn narrowly avoids intersecting the KOS. A total  $\Delta V$ of $2.31$ m/s represents a $35\%$ reduction in propellant usage over the control case in Figure \ref{fig:traj_prescribed}. 

\begin{figure}[h!]
	\centering
	\includegraphics[width=0.75\linewidth]{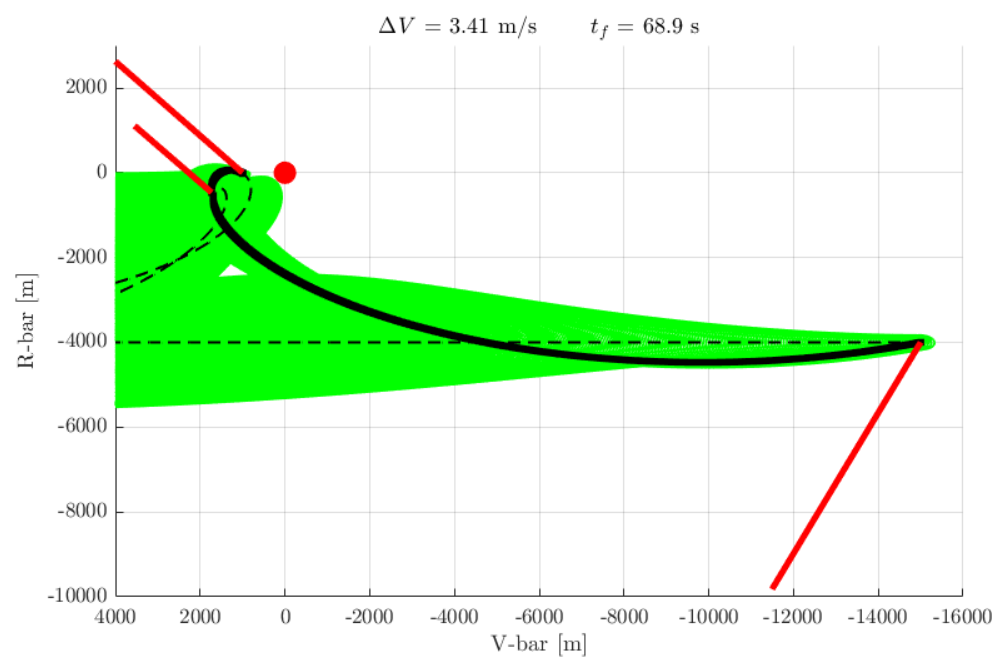}
	\caption{Minimum-time trajectory}
	\label{fig:traj_min_time}
\end{figure}

Figure \ref{fig:traj_min_time} illustrates the results of solving Algorithm 1 identically to how it solved in Figure \ref{fig:traj_min_prop}, but now optimizing for minimum time instead of minimum propellant. Again, the free-drift probability bounds emanating from the penultimate burn very nearly intersect the KOS, indicating that the constraint is active and a limiting factor. The time of flight of 68.9 minutes and $\Delta V$ of $3.41$ m/s represents a $43\%$ reduction in transfer time while using less propellant over the control case. 

%\section{Future work}
\section{Acknowledgements}
The work described in this paper has been privately funded by Blue Origin. The authors would like to thank David Woffinden from NASA JSC for discussions and collaboration through Space Act Agreement No. SAA-EA-20-29343. 

\bibliographystyle{AAS_publication}   % Number the references.
\bibliography{GNSki_RPOD_2022.bib}   % Need to move the references below to a references file

\end{document}